\documentclass[aps,amsfonts,pra,twocolumn,showpacs]{revtex4}
\usepackage{epsfig,amsmath,amssymb,bm,epsf,graphics}
\def\bra#1{\langle#1\vert}
\def\ket#1{\vert#1\rangle}
\def\ketbra#1{\vert#1\rangle\langle#1\vert}

\def\ipr#1#2{\langle#1\vert#2\rangle}

\begin{document}
\title{Exchange symmetry and global entanglement and full separability}
 \author{Tzu-Chieh Wei}
 \affiliation{
    Department of Physics and Astronomy,
University of British Columbia, Vancouver, BC V6T 1Z1, Canada}
\date{\today}
\begin{abstract}
Ichikawa et al. [Phys. Rev. A {\bf 78}, 052105 (2008)] showed that exchange
symmetry gives rise to simple characterization of whether multipartite pure
quantum states being either globally entangled or fully separable.  In this
Brief Report, we provide a simple alternative approach and some extension to
their conclusions.
\end{abstract}
 \maketitle

 Entanglement is one of
many important properties for quantum systems. It is identified as a resource
for many quantum information processing tasks~\cite{NielsenChuang00}. The
characterization and quantification of entanglement in both bipartite and
multipartite settings~\cite{Horodecki4,PlenioVirmani07} has uncovered many
interesting aspects of it, and this has in turn enabled and prompted
investigation of entanglement in many-body systems and connection to quantum
phase transitions~\cite{AmicoFazioOsterlohVedral08,Schollwock05}.

Symmetry usually makes simpler the classification of properties of a
system~\cite{Weyl}.  Recently, Ichikawa et al. employed group-theoretical
method and showed that exchange symmetry gives rise to simple characterization
of multipartite quantum states being either globally entangled or fully
separable~\cite{IchikawaSasakiTsutsuiYonezawa08}.  Here, we provide a simple
alternative approach to their conclusions.

Let us start with a multipartite system comprising $n$ parts, each of which
can have a distinct Hilbert space. Consider a general $n$-partite pure state
(expanded in the local bases $\{|e_{p_i}^{(i)}\rangle\}$):
\begin{equation}
\label{eqn:psi} |\psi\rangle=\sum_{p_1\cdots p_n}\chi_{p_1p_2\cdots p_n}
\ket{e_{p_1}^{(1)}}\otimes\ket{e_{p_2}^{(2)}}\otimes\cdots\otimes\ket{
e_{p_n}^{(n)}}.
\end{equation}
 Let us also give precise
definitions of global entanglement and full separability.\\

\noindent {\bf Definitions.} A state is globally entangled if it remains
entangled across any bi-partition. A state is fully separable if it remains
separable across all bi-partitions. A fully separable pure state is also
called a completely product state and can be expressed in the form
$\ket{\Phi}\equiv\mathop{\otimes}_{i=1}^n|\phi^{(i)}\rangle$.

We note that in the  mixed-state scenario separable states and product states
can mean different things whereas they are the same in the pure-state
scenario. Let us clarify this. A product mixed state across bi-partition A:B
is of the form $\rho=\rho^A\otimes\rho^B$, where $\rho$'s are the
corresponding density matrices. This includes product pure states as a special
case. A completely product mixed state is then of the form
$\rho=\rho^{(1)}\otimes\rho^{(2)}\otimes\dots\otimes\rho^{(n)}$. A separable
mixed state across bi-partition A:B is a state that can be written as
$\rho=\sum_i p_i \rho^A_i\otimes\rho^B_i$, where $p_i\ge0$ and $\sum_i p_i=1$.
A fully separable mixed state is a state that can be written as $\rho=\sum_i
p_i \rho^{(1)}_i\otimes\rho^{(2)}_i\otimes\dots\otimes\rho^{(n)}_i$. Similar
to the pure-state case, we will call a mixed state being globally entangled,
if it is not separable across any bi-partition. For a more refined
classification of mixed states, we refer the readers to Ref.~\cite{Acin01}.
The structure of mixed-state entanglement is much richer, and exotic states,
such as bound entangled states can occur~\cite{Acin01,Bennett99}. We will be
mainly concerned with pure states, but will make a brief comment on
generalization of both results (below) to mixed states at the end. \smallskip

More often than not, the existence of symmetry helps to reduce the difficulty
of the problem and make the solution simpler. For example, when $\ket{\psi}$
is invariant under permutations of parties (namely when the coefficients
$\chi$'s in Eq.~(\ref{eqn:psi}) are invariant under permuting their indices),
simplification arises as to the quantification of their
entanglement~\cite{WeiSeverini09,WeiEricssonGoldbartMunro04,Wei08,HayashiMarkhamMuraoOwariVirmani08,HayashiMarkhamMuraoOwariVirmani09,Robert},
via, e.g., the relative entropy of the
entanglement~\cite{VedralPlenioRippinKnight97}, the geometric
measure~\cite{MiyakeWadati01,BihamNielsenOsborne02,WeiGoldbart03}, or the
Majorana representation~\cite{Markham}, and the characterization of ``exotic''
bound entangled states~\cite{GuhneToth}, as well as proposals for direct
measurements of entanglement~\cite{Enk}. Below, we analyze the global
entanglement and full separability of symmetric states (states that possess
symmetry under permutations), considered earlier by Ichikawa et
al.~\cite{IchikawaSasakiTsutsuiYonezawa08}. They used group theoretical
arguments to
deduce the two following main results:\smallskip \\
\noindent {\bf Result 1}: Symmetric states are either globally entangled or
fully separable with all the constituent systems having identical states,
whereas antisymmetric states are  globally entangled. \smallskip\\
\noindent {\bf Result 2}: No completely product states can be orthogonal to
all symmetric states and symmetrization of a completely product state gives
rise to a globally entangled state unless the original product state is
symmetric (namely, with all the constituent systems having identical states).

We shall provide an alternative approach to these results.

\noindent {\bf Proof of Result 1}:\\
 Let us, for the sake of argument, imagine that the parties are arranged on a
circle and consider that the state has the symmetry
\begin{equation}
{T}\ket{\psi}=e^{i \theta}\ket{\psi},
\end{equation}
where ${T}$ is the periodic translation on party labels: $1\rightarrow 2$,
$2\rightarrow 3$, ..., $n\rightarrow 1$. We claim that if $\ket{\psi}$ is
separable under any bi-partition of the form $\{1,2,..,k: k+1,k+2,..,n\}$ then
$\ket{\psi}$ must be fully separable. The proof is as follows. The
bi-separability implies that
\begin{equation}
\ket{\psi}=\ket{\phi^A}_{1,2,..,k}\otimes\ket{\phi^B}_{k+1,k+2,..,n}.
\end{equation}
The translation symmetry implies that
$\ket{\phi}_{1,2,..,k}\otimes\ket{\phi}_{k+1,k+2,..,n}$ is also separable
under the partition ${2,..,k+1: k+2,k+3,..,n,1}$ and hence
\begin{equation}
\ket{\psi}=\ket{\phi^{A_1}}_{1}\otimes\ket{\phi^{A_2}}_{2,..,k}\otimes\ket{\phi^{B_1}}_{k+1}
\otimes \ket{\phi^{B_2}}_{k+2,..,n}.
\end{equation}
The argument holds regardless of the value $\theta$, as what matters is the
separability. Continuing this argument, we arrive at that $\ket{\psi}$ must be
fully separable. We have thus shown that states that possess translational
symmetry are either globally entangled or fully separable, extending results
of Ichikawa et al.

A periodic translation operator is also an element in the permutation group
and hence, if a permutation invariant state is separable under any
bi-partition (and by permutation the partition can be made to be $\{1,2,..,k:
k+1,k+2,..,n\}$), it must be fully separable. When a permutation invariant
state is separable, its constituent systems must possess identical states (up
to irrelevant global phases). Therefore, we concludes that a permutation
invariant state (as a special case of translation invariant states) is either
globally entangled (i.e., it cannot be separable across any bi-partition) or
fully separable.

We note that the argument can be applied to the totally antisymmetric state as
well, as it also satisfies ${T}\ket{\psi}=\pm \ket{\psi}$. But it cannot be
separable, as this implies complete separability, which cannot induce a sign
change under any permutation. Thus, an totally antisymmetric state is
entangled, and hence globally entangled. Furthermore, it can also be applied
to the case of braid group ${\cal B}_n$ when the state $\ket{\psi}$ satisfies
\begin{equation}
\pi_\sigma \ket{\psi}=e^{i\theta} \ket{\psi}, \ \ \forall\, \pi_\sigma\in
{\cal B}_n,
\end{equation}
as a periodic translation operator can be constructed from elements in the
braid group. For all states $\ket{\psi}$ satisfying the above symmetry with
$e^{i\theta}\ne 1$, they are necessarily globally entangled. Hence, we have
provide alternate proof of Result 1, which was originally proven in
Ref.~\cite{IchikawaSasakiTsutsuiYonezawa08} using group-theoretical arguments.
{\hfill $\Box$}\medskip

\noindent {\bf Proof of Result 2}:\\
Next, we shall derive Result 2. We first show that for any product state
$\ket{\Phi}=\ket{\phi_1}\otimes\ket{\phi_2}\dots\otimes\ket{\phi_n}$, the
symmetrization yields a nonzero state
\begin{equation}
\label{eqn:PhiS} \ket{\Phi_S}= \frac{c}{n!}\sum_{\sigma\in{\cal S}_n}\sigma
\ket{\Phi},
\end{equation}
i.e., $\ipr{\Phi_S}{\Phi_S}>0$, where $c\ne 0$ is a normalization constant and
${\cal S}_n$ denotes the permutation group for $n$ objects. It is
straightforward to see that
\begin{equation}
\ipr{\Phi_S}{\Phi_S}=\frac{|c|^2}{n!}{\rm Perm}(\ipr{\phi_i}{\phi_j}),
\end{equation}
where Perm denotes the permanent of a matrix. As the matrix
$a_{ij}\equiv\ipr{\phi_i}{\phi_j}$ is positive semi-definite and $a_{ii}=1$,
from the results of Marcus~\cite{Marcus63}, we see that the permanent of
matrix $a_{ij}$ is nonzero, and in particular
\begin{equation}
1\le {\rm Perm}(a_{ij})\le n!,
\end{equation}
where the first inequality becomes equality when $a_{ij}=\delta_{ij}$ and the
second inequality becomes equality when $a_{ij}$ is rank-one, i.e.,
$\ket{\phi_i}$'s are identical up to a phase factor. What we have just shown
is that the symmetrization of a state composed of direct product of
single-particle states always yields a valid state, which is an implicit
assumption in discussing a bosonic state. Conversely, if
$\ipr{\Phi_S}{\Phi_S}=0$, this implies that $\ipr{\Phi}{\Phi}=0$ as well.

Now, suppose there exists a completely product state $\ket{\Phi}$ that is
orthogonal to all permutation invariant states. As the inner product of
$\ket{\Phi}$ with any such symmetric state $\ket{\chi}$ is invariant under
permuting parties in $\ket{\Phi}$, this means that its symmetrized state
$\ket{\Phi_S}$ must be orthogonal to all permutation invariant states, hence
including itself! This leads to contradiction. Therefore, there cannot exist a
completely product state that is orthogonal to all permutation invariant
states.

Furthermore, from Result 1 the symmetrized state $\ket{\Phi_S}$ must be either
globally entangled or fully separable. In the latter case,
$\ket{\Phi_S}=\ket{\phi}^{\otimes n}$ (up to normalization and a global
phase), we want to show that this implies that the original state $\ket{\Phi}$
must be uniquely of the form $\ket{\Phi}=\ket{\phi}^{\otimes n}$ up to a
global phase factor. A consequence of $\ket{\Phi_S}$ being a product state is
that its reduced density matrix after tracing over $k=1,...,n-1$ is still
pure. We shall consider tracing over $(n-1)$ parties. Now we can rewrite
$\ket{\Phi_S}$ as

\begin{eqnarray}
\ket{\Phi_s} &\equiv&\frac{c}{n!}\sum_{i=1}^n\ket{\phi_i}_A\ket{\psi_i}_B,
\end{eqnarray}
where the $(n-1)$-partite state $\ket{\psi_k}$ that is associated with
one-partite state $\ket{\phi_k}$ is of the form:
\begin{eqnarray}
\ket{\psi_k}\equiv \sum_{\tilde\sigma\in {\cal S}_{n\!-\!1}} \tilde\sigma
\ket{\phi_{k+1}}...\ket{\phi_n}\ket{\phi_1}...\ket{\phi_{k-1}}.
\end{eqnarray}
Tracing over $(n-1)$ parties (namely $B$), we obtain
\begin{eqnarray}
\rho={\rm
Tr}_{2...n}(\ketbra{\Phi_S})=\frac{|c|^2}{(n!)^2}\sum_{i,j}\ket{\phi_i}\bra{\phi_j}\ipr{\psi_j}{\psi_i}.
\end{eqnarray}
Since $\ipr{\psi_j}{\psi_i}$ is positive semi-definite we can diagonalize it:
\begin{equation}
\ipr{\psi_j}{\psi_i}=\sum_k \lambda_k U_{jk} U_{ik}^*,
\end{equation}
where $U$ is unitary and $\lambda_k\ge 0$. This means that
\begin{eqnarray}
\rho=\frac{|c|^2}{(n!)^2}\sum_{k}\lambda_k\ketbra{\alpha_k},
\end{eqnarray}
where
\begin{equation}
\ket{\alpha_k}\equiv \sum_{i=1}^n U_{ik}^* \ket{\phi_i}.
\end{equation}
But we also have that $\rho=\ketbra{\phi}$, this means that
$\ket{\alpha_k}\sim \ket{\phi}$, which in turn (by inverting the above
equation) gives
\begin{equation}
\ket{\phi_j}= \sum_{k=1}^n U_{jk}\ket{\alpha_k}\sim \ket{\phi}.
\end{equation}
Thus, we have $\ket{\phi_j}=\ket{\phi}$ up to a global phase and
$\ket{\Phi}=\ket{\phi^{\otimes n}}$. Hence, Result 2 is proved. {\hfill
$\Box$}\medskip

Let us conclude with some comments on mixed states. A generalization of Result
1 would be: mixed states that reside in the symmetric subspace (i.e., those
states that are composed of mixture of permutation invariant pure states) are
either globally entangled or fully separable. This is indeed the case. If such
a symmetric mixed state is separable with respect to some bi-partition, this
implies that for certain decomposition all the pure symmetric states in the
mixture are in a product form with respect to this bi-partition. Then by
reasoning in Result 1, this means that they are completely product states.
Hence, the corresponding mixed state is fully separable. A partial
generalization of Result 2 would be: No fully separable mixed states can be
orthogonal to all symmetric mixed states (orthogonal in the sense of this
``inner product'': ${\rm Tr}(\rho_1^\dagger\rho_2)=0$). This is also correct,
as all symmetric mixed states are composed of mixture of pure symmetric
states, this reduces to showing no fully separable mixed states can be
orthogonal to all symmetric pure states. As a fully separable state $\rho$ is
composed of completely product states $\ket{\phi_i}$: $\rho=\sum_i
p_i\ketbra{\phi_i}$, the inner product with a symmetric state $\ket{\psi_S}$
becomes $\sum_i p_i |\ipr{\phi_i}{\psi_S}|^2$. The expression is zero if and
only if $\ipr{\phi_i}{\phi_S}=0$ for all $i$ (with $p_i> 0$). If this holds
for all symmetric states $\ket{\psi_S}$, it will lead to the same
contradiction that $\ket{\phi_i}=0$ in Result 2. Hence, the generalization to
mixed states is also correct. However, a generalization of the second part
regarding symmetrization of fully separable mixed states does not lead to any
interesting outcome, namely, it will not make fully separable mixed states
become globally entangled. Because we define the symmetrization (sum over all
permutations) at the level of density matrices, a fully separable state
remains fully separable under symmetrization.

 \noindent {\bf
Acknowledgment.} This work was supported by NSERC and MITACS.


\begin{thebibliography}{000}
\bibitem{NielsenChuang00}
M. Nielsen and I. Chuang, {\sl Quantum Computation and Quantum Information\/}
(Cambridge University Press, Cambridge, 2000).
\bibitem{Horodecki4}
R. Horodecki, P. Horodecki, M. Horodecki, and K. Horodecki, Rev. Mod. Phys.
{\bf 81}, 865 (2009).
\bibitem{PlenioVirmani07}
 M. B. Plenio and S. Virmani, Quant. Inf. Comp. {\bf
7}, 1 (2007).
\bibitem{AmicoFazioOsterlohVedral08}
 L. Amico, R. Fazio, A. Osterloh, and V. Vedral. Rev.
Mod. Phys. {\bf 80}, 517 (2008).
\bibitem{Schollwock05}
 U. Schollw\"ock, Rev. Mod. Phys. {\bf 77}, 259
(2005).
\bibitem{Weyl}
H. Weyl, {\sl Symmetry\/} (Princeton University Press, Princeton, 1952).
\bibitem{IchikawaSasakiTsutsuiYonezawa08}
T. Ichikawa, T. Sasaki, I. Tsutsui, and N. Yonezawa, Phys. Rev. A {\bf  78},
052105 (2008).
\bibitem{Acin01} A. Acin, D. Bru\ss, M. Lewenstein, and A. Sanpera, Phys. Rev.
Lett. {\bf 87}, 040401 (2001).
\bibitem{Bennett99}
C. H. Bennett, D. P. DiVincenzo, T. Mor, P. W. Shor, J. A. Smolin, and B. M.
Terhal, Phys. Rev. Lett. {\bf 82}, 5385 (1999).
\bibitem{HayashiMarkhamMuraoOwariVirmani08}
M. Hayashi, D. Markham, M. Murao, M. Owari, and S. Virmani, Phys. Rev. A {\bf
77}, 012104 (2008).
\bibitem{HayashiMarkhamMuraoOwariVirmani09}
M. Hayashi, D. Markham, M. Murao, M. Owari, and S. Virmani, J. Math. Phys.
{\bf 50}, 122104 (2009); also in arXiv:0905.0010.
\bibitem{WeiEricssonGoldbartMunro04}
T.-C. Wei, M. Ericsson, P. M. Goldbart, and W. J. Munro, Quantum Inf. Comput.
{\bf 4}, 252 (2004).
\bibitem{Wei08}
T.-C. Wei, Phys. Rev. A {\bf 78}, 012327 (2008).
\bibitem{Robert}
R. H\"ubener, M. Kleinmann, T.-C. Wei, C. Gonz\'alez-Guill\'en and O. G\"uhne,
 Phys. Rev. A {\bf 80}, 032324 (2009).
\bibitem{WeiSeverini09}
T.-C. Wei and S. Severini, arXiv:0905.0012.
\bibitem{VedralPlenioRippinKnight97}
V. Vedral, M. B. Plenio, M. A. Rippin, and P. L. Knight, Phys. Rev. Lett. {\bf
78}, 2275 (1997).
\bibitem{MiyakeWadati01}
A. Miyake and M. Wadati, Phys. Rev. A {\bf 64}, 042317 (2001).
\bibitem{BihamNielsenOsborne02}
O. Biham, M. A. Nielsen, and T. J. Osborne, Phys. Rev. A {\bf 65}, 062312
(2002).
\bibitem{WeiGoldbart03}
T.-C. Wei and P. M. Goldbart, Phys. Rev. A {\bf 68}, 042307 (2003).
\bibitem{Markham}
D. J. H. Markham, arXiv:1001.0343.
\bibitem{GuhneToth}
G. T\'oth and O. G\"uhne, Phys. Rev. Lett. {\bf 102}, 170503 (2009).
\bibitem{Enk}
S. J. van Enk, Phys. Rev. Lett. {\bf 102}, 190503 (2009).
\bibitem{Marcus63}
M. Marcus, Bull. Amer. Math. Soc. {\bf 69}, 494 (1963).

\end{thebibliography}
\end{document}